\documentclass{natureprintstyle}
\usepackage{ifpdf} 
\ifpdf
  \pdfoutput=1  
 \usepackage[pdftex]{graphicx}         
  \usepackage{pdftricks}
  \usepackage{epstopdf}
  \usepackage[colorlinks, bookmarks, breaklinks, pdftitle={},pdfauthor={Remco C. E. van den Bosch}]{hyperref}
   
  \hypersetup{linkcolor=black,citecolor=black,filecolor=black,urlcolor=blue}       
  \usepackage[kerning,final]{microtype}  
\else                                
  \usepackage[dvips]{graphicx} 
\fi

\usepackage{amssymb}
\usepackage{relsize}

\newcommand{\apj}{Astrophys. J.}           
\newcommand{\apjl}{Astrophys. J.}           
\newcommand{\mnras}{Mon. Not. R. Astron. Soc.}       
\newcommand{\nat}{Nature}
\newcommand{\aap}{Astron. Astrophys.}

\newcommand{\aj}{Astron. J.}
\newcommand{\pasp}{Pubbl. Astron. Soc. Pacific}
\newcommand{\apjs}{Astrophys. J. Suppl.}           

\newcommand{\spie}{Proc. Soc. Photo-Opt. Instrum. Eng.}

\def\kms{km\,s$^{-1}$}
\def\dgr{$^\circ$}
\def\Mbh{$M_\bullet$}
\def\Msun{M$_\odot$}

\def\MLsunV{M$_\odot$/L${_{\odot,V}}$}

\def\arcs{\hbox{$^{\prime\prime}$}}
\def\arcsec{\hbox{$^{\prime\prime}$}} 
\def\arcm{\hbox{$^{\prime}$}}
\def\Msigma{\Mbh --\relsize{-1}$\sigma$\relsize{+1}}

\def\Wfpc{\textsc{wfpc}}
\def\Wfpc2{\textsc{wfpc\relsize{-1}2\relsize{+1}}}

\title{An over-massive black hole in the compact lenticular galaxy NGC\,1277}

\ifpdf
\author{ 
\href{http://mpia.de/~bosch/}                  {Remco C. E. van den Bosch}$^{1,2}$, 
\href{http://www.as.utexas.edu/~gebhardt/}     {Karl Gebhardt}$^{2}$,  
\href{http://dept.astro.lsa.umich.edu/~kayhan/}{Kayhan G\"ultekin}$^{3}$, 
\href{http://mpia.de/~glenn/}                  {Glenn~van~de~Ven}$^{1}$, 
\href{http://mpia.de/~vdwel/}                  {Arjen~van~der~Wel}$^{1}$ \& 
\href{http://www.as.utexas.edu/astronomy/people/people.html?u=55}{Jonelle~L.~Walsh}$^{2}$ \\
} 
\else
\author{Remco C. E. van den Bosch$^{1,2}$, Karl Gebhardt$^{2}$,  Kayhan G\"ultekin$^{3}$, Glenn~van~de~Ven$^{1}$, Arjen~van~der~Wel$^{1}$ \& Jonelle~L.~Walsh$^{2}$ \\
}
\fi

\begin{document}

\maketitle 

\begin{abstract} 

All massive galaxies likely have supermassive black holes at their centers, and the masses of the black holes are known to correlate with properties of the host galaxy bulge component\cite{2009ApJ...698..198G}. Several explanations have been proposed for the existence of these locally-established empirical relationships; they include the non-causal, statistical process of galaxy-galaxy merging\cite{2011ApJ...734...92J}, direct feedback between the black hole and its host galaxy\cite{1999MNRAS.308L..39F}, or galaxy-galaxy merging and the subsequent violent relaxation and dissipation\cite{2011Natur.469..374K}. The empirical scaling relations are thus important for distinguishing between various theoretical models of galaxy evolution\cite{2006MNRAS.365...11C,2008MNRAS.391..481S}, and they further form the basis for all black hole mass measurements at large distances. In particular, observations have shown that the mass of the black hole is typically 0.1\% of the stellar bulge mass of the galaxy\cite{2004ApJ...604L..89H, 2011MNRAS.413.1479S}. The small galaxy NGC\,4486B currently has the largest published fraction of its mass in a black hole at 11 per cent \cite{1998AJ....115.2285M, 2009ApJ...698..198G}. Here we report observations of the stellar kinematics of NGC 1277, which is a compact, disky galaxy with a mass of {\boldmath$1.2 \times 10^{11}$} \Msun. From the data, we determine that the mass of the central black hole is {\boldmath $1.7\times10^{10}$} \Msun, or 59\% its bulge mass. Five other compact galaxies have properties similar to NGC 1277 and therefore may also contain over-sized black holes. It is not yet known if these galaxies represent a tail of a distribution, or if disk-dominated galaxies fail to follow the normal black hole mass scaling relations\cite{2010MNRAS.403..646N, 2011Natur.469..374K}.

\end{abstract} 

\noindent \let\thefootnote\relax\footnote{\textbf{{\small Received 4 May; accepted 13 September; published November 29 2012}}}
\noindent \let\thefootnote\relax\footnote{\noindent \begin{affiliations}
\ifpdf
\item \href{http://mpia.de/}{Max-Planck Institut f\"ur Astronomie, Heidelberg, Germany}
\item \href{http://www.as.utexas.edu/}{Department of Astronomy, University of Texas, Austin, USA.}
\item \href{http://dept.astro.lsa.umich.edu/}{Department of Astronomy, University of Michigan, Ann Arbor, USA.}
\else
\item Max-Planck Institut f\"ur Astronomie, K\"onigstuhl 17, D-69117 Heidelberg, Germany
\item Department of Astronomy, University of Texas, Austin, Texas 78712, USA.
\item Department of Astronomy, University of Michigan, Ann Arbor, Michigan 48109, USA.
\fi
\end{affiliations}}

Direct black hole mass measurements often rely on obtaining spatially resolved stellar or gas kinematics within the region over which the black hole dominates the gravitational potential, the so-called sphere-of-influence. We have obtained long-slit spectroscopy  of 700 nearby galaxies with the Marcario Low Resolution Spectrograph\cite{1998SPIE.3355..375H} on the Hobby-Eberly Telescope to find suitable targets for direct black hole mass measurements (see SI). As shown in Table~\ref{tab:properties}, six of these galaxies have very peculiar properties; they have velocity  dispersion $\sigma>350$\,\kms\ and half-light radii $R_e < 3$\,kpc. It is unusual for such small galaxies to have such large dispersions, signifying an unusually high central mass concentration: a simple virial mass estimate indicates that the central 200 parsec contains more than 10 billion solar masses which is one hundred times more than typical for galaxies of the same size. 

\begin{figure*}
\centering 
\includegraphics[width=0.8\textwidth]{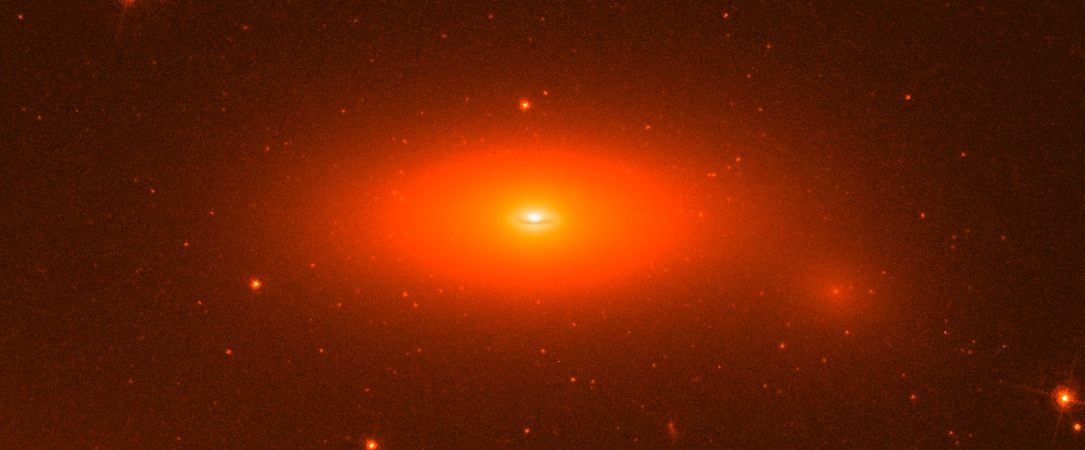}
	\caption{\textbf{Optical \emph{Hubble Space Telescope} image of the compact lenticular galaxy NGC\,1277.} The image is shown with a logarithmic stretch and is $19\times8$ kpc, with north pointing up and east to the left. In this high resolution optical image the galaxy has a half-light radius of 1\,kpc, is strongly flattened, and disky. It is immediately clear that a superposition of multiple galaxies does not explain the high velocity dispersion. NGC\,1277 has a small regular nuclear dust disk with an apparent axis ratio of only 0.3, which indicates that we see the galaxy close to edge-on. Through a multi-component fit{\protect \cite{2002AJ....124..266P}} to the \emph{HST} image, we identify the inner component, with a half-light-radius of 0.3 kpc and a Sersic index of $n\simeq1$ as a pseudo-bulge, that contains 24\% of the light. For the dynamical modelling we construct a three-dimensional luminous mass model of the stars by de-projecting the two-dimensional light model from the \emph{HST} image. Then, the gravitational potential is inferred from the combined luminous mass, black hole mass, and dark matter halo. In this potential representative orbits are integrated numerically, while keeping track of the paths and orbital velocities of each orbit. We then create a reconstruction of the galaxy by assigning each orbit an amount of light, so that the model recreates the total light distribution, while simultaneously fitting the long-slit stellar kinematics observed with the Hobby-Eberly Telescope (Fig.~\ref{fig:NGC1277slit}). The models include the effect of the Earth's atmosphere and the telescope optics without any \emph{a priori} assumption on the orbital configuration. See SI.
	See an animation of the orbits in this galaxy at 
	\ifpdf
	\href{http://mpia.de/~bosch/blackholes.html}{\url{http://mpia.de/~bosch/blackholes.html}}
	\else
	\url{http://mpia.de/~bosch/blackholes.html}
	\fi
	}
	\label{fig:NGC1277pretty}
\end{figure*}

\begin{figure}[t]
	\centering
		\includegraphics[width=8.7cm]{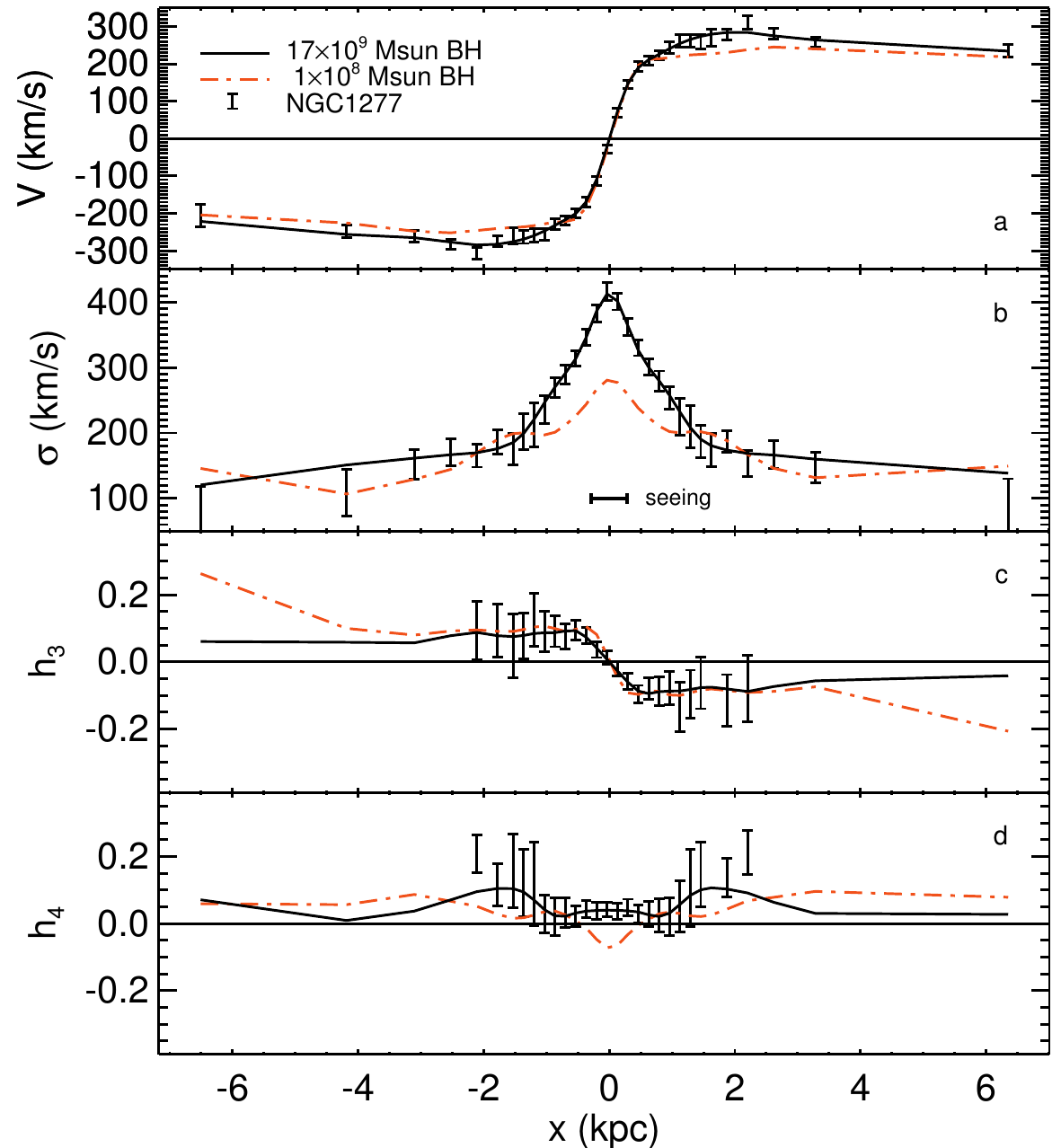}
	\caption{\textbf{Line-of-sight stellar kinematics of NGC\,1277.} The stellar kinematics as observed with the Marcario Low Resolution Spectrograph\cite{1998SPIE.3355..375H} on the Hobby-Eberly Telescope, shown with the $1\sigma$ error bars, were measured{\protect\cite{2004PASP..116..138C}} at 31 locations along the major axis of NGC\,1277 (See SI.). Panels a,b,c and d, from top to bottom, show the  mean velocity, velocity dispersion, and higher-order Gauss-Hermite velocity moments{\protect\cite{1993ApJ...407..525V}} $h_3$ and $h_4$, representing skewness and kurtosis, respectively. The kinematics show a remarkably flat rotation curve and a dispersion profile that strongly peaks toward the center. The best-fit Schwarszchild model (black line) has a $17\times10^{9}$ \Msun\ black hole. The relation between black-hole mass and host luminosity predicts a $10^8$ \Msun\ black hole, but the corresponding model (red dot-dashed line) does not fit the data at all. The telescope resolution (seeing 1.6\arcs\ FWHM) is indicated in panel b and is sufficient to resolve the Sphere-of-Influence of the black hole. }
	\label{fig:NGC1277slit}
\end{figure}

Black hole masses can be measured directly by fitting self-consistent Schwarzschild models\cite{1979ApJ...232..236S} to spatially resolved spectroscopy and high resolution imaging. Incidentally, archival Hubble Space Telescope (\emph{HST}) imaging is available for one of these six dense galaxies, NGC\,1277. Based on the \emph{HST} imaging (Figure~\ref{fig:NGC1277pretty}) and the stellar kinematics (Figure~\ref{fig:NGC1277slit}) we constructed 600\,000 orbit-based models using iterative refinement to search parameter space\cite{2008MNRAS.385..647V, 2010MNRAS.401.1770V}. The best-fit model is then found by marginalizing over all parameters: the stellar mass-to-light ratio, the black-hole mass, and the dark halo\cite{1996ApJ...462..563N}. The confidence intervals are determined with the goodness-of-fit statistic $\chi^2$. We measure a black-hole mass of 17$\pm3\times 10^9$ \Msun\ and total stellar mass of $1.2\pm0.4\times10^{11}$ \Msun, with $1\sigma$ confidence intervals based on $\Delta \chi^2=1$ after marginalizing over the dark halo parameters (see SI.). The black hole in NGC\,1277 is one of the most massive black holes to be dynamically confirmed, and moreover has a mass fraction of 14 per cent of the total stellar mass in this galaxy.

\begin{figure}
	\centering
		\includegraphics[scale=0.9]{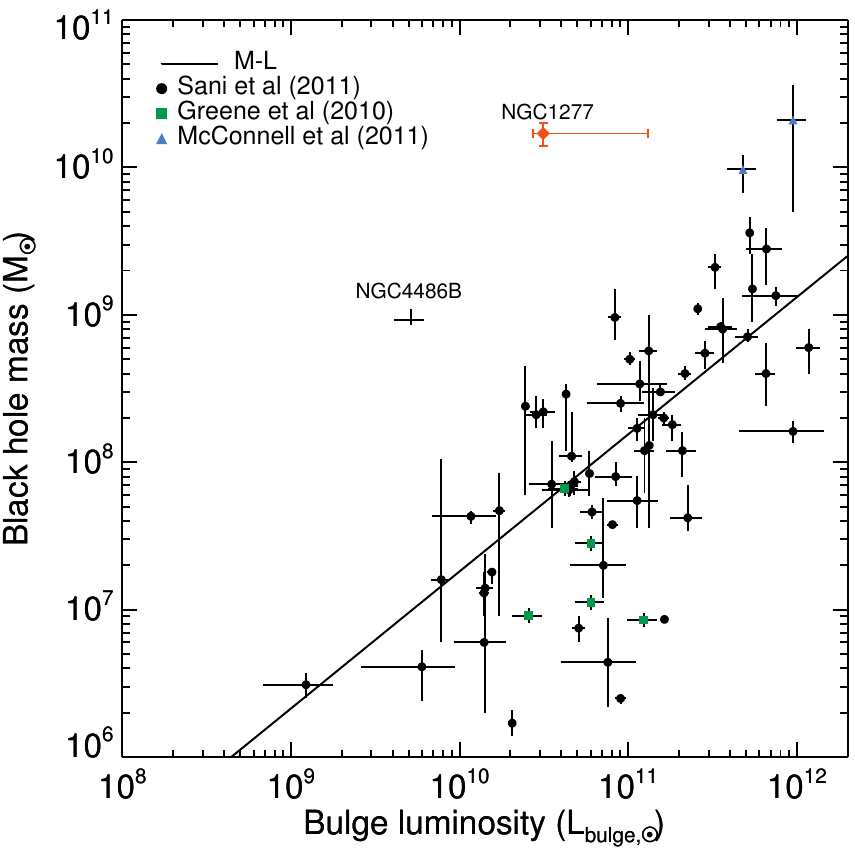}
	\caption{\textbf{The correlation of black hole mass and near-infrared bulge luminosity.} The black line shows the black-hole-mass--bulge-luminosity relation{\protect \cite{2011MNRAS.413.1479S}} for galaxies with a directly measured black hole mass. NGC~1277 is a significant positive outlier. In addition to the galaxies (black dots) to which the relation has been fitted{\protect \cite{2011MNRAS.413.1479S}}, nine black hole masses{\protect\cite{1998AJ....115.2285M, 2010ApJ...721...26G,2011Natur.480..215M}} have been added with 2MASS K-band bulge luminosities. The error bars denote $1\sigma$ uncertainties, except for the NGC\,1277 bulge luminosity, where we use its total luminosity as a conservative upper limit.}
 
 \label{fig:scaleML} \end{figure}

No galaxy has previously been seen with such a large ratio of black-hole mass to stellar mass. Due to the strong disk-like rotation (Fig.~\ref{fig:NGC1277slit}) and the lack of an unambiguous bulge in NGC\,1277 (Fig.~\ref{fig:NGC1277pretty}), it is difficult to place it on the relation between black-hole mass and bulge luminosity. The central pseudo-bulge contains 24\% of the light (see Figure~\ref{fig:NGC1277pretty}) and  black hole-to-bulge mass fraction is 59 per cent. As shown in Fig~\ref{fig:scaleML}, NGC\,1277 is a significant outlier from the black-hole-bulge relation by two orders of magnitude. At a fixed bulge luminosity of $3$--$10\times10^{10}$ L$_{K,\odot}$, dynamical black-hole-mass measurements now range over four orders of magnitude, from $10^6$ to $10^{10}$ \Msun, showing that (pseudo-)bulge luminosity is not a good predictor for black-hole mass.

We now place NGC1277 on the relation between black-hole mass and velocity dispersion (\Msigma). The average velocity dispersion inside the half-light radius  (2.8\arcs) and outside the Sphere-of-Influence (1.6\arcs) for NGC\,1277 is $\sigma=333$\,\kms, based on a reconstruction of the best-fit orbit-based model. For this $\sigma$, the most recent inferred relation\cite{2011Natur.480..215M} predicts a black-hole mass of $2.4\times10^9$\,\Msun, so the measured value is almost one magnitude higher, or a $2.1\sigma$ outlier relative to the intrinsic scatter in the \Msigma\ relation\cite{2009ApJ...698..198G}. 

Besides NGC1277, NGC\,4486B\cite{1998AJ....115.2285M} and Henize\,2-10\cite{2011Natur.470...66R} are known to lie significantly above the relations. At the same time, at least 3 galaxies are known to lie significantly below the relations, too\cite{2001Sci...293.1116M, 2010MNRAS.403..646N, 2010ApJ...721...26G}. We do not yet know if these over-massive and under-massive black holes are just the tail of a relatively narrow distribution of joint black-hole--galaxy properties, or if they demonstrate non-universality. Only through more black hole measurements, including those in the other five compact galaxies with high velocity dispersions, we will be able to establish the cause of the black-hole--galaxy connection.

A stellar population analysis of NGC1277\cite{2005MNRAS.358..363C} indicates there are only old ($\gtrsim 8$\,Gyr) stars present and there has not been any recent star formation. The black hole must thus have been in place since that time, as a scenario where black hole accretion happens without significant star formation or the formation of a (classical) bulge is highly unlikely. Furthermore, there is no strong evidence that NGC1277 has been tidally stripped, as its isophotes are extremely regular and disky and appears to have a normal dark matter halo inferred from the dynamical model, and at large radii the rotation curve is flat out to five times the half-light radius.           
 
Whereas the six compact galaxies presented in Table~\ref{tab:properties} are unusual and rare in the present-day universe, they are, interestingly enough, quantitatively similar to the typical red, passive, galaxies at much earlier times (at redshifts $z\sim 2$): those are also found, on average, to be smaller than similarly massive galaxies in the present-day universe\cite{2008ApJ...677L...5V}, possibly possess high velocity dispersions\cite{2009Natur.460..717V}, and generally have a disk-like structure\cite{2011ApJ...730...38V}.  Perhaps the compact systems we found are local analogues of these high-redshift galaxies.

\section*{\relsize{-1}{\smaller Received 4 May; accepted 13 September; published November 29 2012}\relsize{+1}}

\begin{table}
	\begin{center}
	\relsize{-1}
	\begin{tabular}{lcccccc}   
		\hline
		\hline 
Object & Distance & $\sigma_c$  & $R_{e,K}$  & Luminosity & $\epsilon$  \\ 
          & (Mpc)   & (\kms) & (kpc) &log(L$_K$) & ($1-b/a$)\\ 
          & (1)   & (2)&   (3) & (4) & (5)   \\
\hline     
  ARK90   & 131  &    392$\pm$4 & 1.6&11.2& 0.7   \\ 
 NGC1270  &  69  &    393$\pm$3 & 1.8&11.2& 0.8   \\ 
 NGC1277  &  73  &    403$\pm$4 & 1.6&11.1& 0.5   \\ 
UGC1859   &  82  &    362$\pm$4 & 2.0&11.2& 0.6   \\ 
UGC2698   &  89  &    397$\pm$3 & 2.7&11.4& 0.7   \\ 
 MRK1216  &  94  &    354$\pm$4 & 1.9&11.2& 0.6   \\ 
\hline
	\end{tabular}
	\end{center}%
	\caption{\textbf{Global properties of the six compact, high-dispersion galaxies.} 	The six galaxies presented here were observed with the Marcario Low Resolution Spectrograph{\protect\cite{1998SPIE.3355..375H}} on the Hobby-Eberly Telescope as part of a large survey program. We targeted galaxies from
	 the Two Micron All Sky Survey (2MASS) extended source catalog{\protect\cite{2000AJ....119.2498J}} that are expected to have the largest Sphere-of-Influence. Our predictions of the Sphere-of-Influence assume that the galaxies follow the relation between black-hole mass and host galaxy velocity dispersion{\protect\cite{2009ApJ...698..198G}}. For those 2MASS galaxies without a velocity dispersion value from the literature, we used an estimate based on the Fundamental Plane relation between galaxy size, surface brightness and velocity dispersion{\protect\cite{1998AJ....116.1591P}}. See the SI for for more information on the survey. The columns show the near-infrared properties (1) Distance from Hubble flow. (2) Stellar velocity dispersion extracted from a central aperture. (3,4,5) 2MASS{\protect \cite{2000AJ....119.2498J}} half-light radius, total luminosity and apparent ellipticity. 
	}
\label{tab:properties}
\end{table}

\begin{addendum}  
	
\item[Acknowledgements] KG and JW are supported by the National Science Foundation (NSF-0908639, AST-1102845). KG{\"u} acknowledges support provided by National Aeronautics Space Administration (GO0-11151X, G02-13111X) and Space Telescope Science Institute (HST-GO-12557.01-A). The Hobby-Eberly Telescope is a joint project of the University of Texas at Austin, the Pennsylvania State University, Ludwig-Maximilians-Universit\"at M\"unchen, and Georg-August-Universit\"at G\"ottingen. The Hobby-Eberly Telescope is named in honor of its principal benefactors, William~P.~Hobby and Robert~E.~Eberly.

\item[Author Contributions] RvdB designed the survey and carried out the data analysis and the modeling. RvdB and GvdV wrote the manuscript. AvdW carried out the image analysis. All authors contributed to the interpretation of the observations and the writing of the paper.

\item[Author Information] Correspondence and requests for materials should be addressed to RvdB (\ifpdf \href{mailto:bosch@mpia.de}{    
bosch@mpia.de}). 
\else
[bosch@mpia.de]).
\fi
  
\item[Supplementary Information] begins on the next page.  
            
\end{addendum}

\clearpage
\newpage
\renewcommand{\figurename}{Supplemental Figure}
\renewcommand{\tablename}{Supplemental Table}
\setcounter{figure}{0}
\setcounter{table}{0}

\begin{center}
\title  {\huge \bf{Supplementary Information}}
\end{center}

This Supplementary Information consists of four parts.
The first part describes the HET Massive Galaxy Survey, and is followed by a description of the stellar kinematics. The third section shows the analysis of the \emph{HST} photometry, and finally the results from the dynamical models are described in the last section.   

Throughout we adopt a Hubble constant $H_0=70$\,\kms Mpc$^{-1}$, a cosmological matter fraction of $\Omega_m=0.3$, an angular diameter distance for NGC\,1277 of 73\,Mpc, and a luminosity distance modulus of $M-m=34.39$. At this distance, 1\arcs\ corresponds to 353 pc.

\section{HET Massive Galaxy Survey\\} 
\label{sec:het_massive_galaxy_survey}

Most massive nearby galaxies do not have a stellar velocity dispersion $\sigma$ measured, and even if they do, the amalgamated literature is often unreliable. The HET Massive Galaxy Survey is designed to remedy this and provide a census of the nearby massive galaxies. This multi-trimester program on the Hobby-Eberly Telescope\cite{1998SPIE.3352...34R} (HET) is targeting the nearest galaxies with the highest expected stellar velocity dispersions with the Marcario Low Resolution Spectrograph\cite{1998SPIE.3355..375H} (LRS). The survey is still ongoing, and thus we only present a brief summary here.

The targets are selected from a master catalog, which is made up of all 13000 sources brighter than $m_{Ks}<11$ in the 2MASS extended source catalog\cite{2000AJ....119.2498J} and are visible from the HET location ($-10<\delta<74$\dgr). This catalog is cross-matched with redshift-based distances from the NASA Extragalactic Database (NED) and the 2MASS redshift survey\cite{2012ApJS..199...26H} as well as redshift-independent distances from NED-1D. Literature stellar velocity dispersions have been obtained from HyperLeda\cite{2003A&A...412...45P} and SDSS NYU-VAGC\cite{2005AJ....129.2562B,2008ApJS..175..297A}.  The Fundamental Plane relation between galaxy size, surface brightness, and velocity dispersion is used to predict the velocity dispersion when no literature value exists. With the combination of the near infrared fundamental plane parameters\cite{1998AJ....116.1591P}, luminosities and sizes from 2MASS, and foreground galactic extinction corrections, we found that the velocity dispersion can be predicted with an accuracy of 35 \kms. From this master-catalog the survey targets those galaxies with the largest Sphere-of-Influence $R_{SOI} =  \frac{G M_\bullet }{ D \sigma^2 }$, given their distance $D$ and black hole mass $M_\bullet$ as predicted from the  \Msigma\ relation\cite{2009ApJ...698..198G}. No criteria on galaxy type are applied. 

\section{Long-Slit Stellar Kinematics\\}

The LRS is a straightforward optical long slit spectrograph with a slit length of 4\arcm\ and a 3k$\times$1k CCD. We used the g2 grating (covering 4200-7400\,\AA), 2$\times$2 binning, and a slit width of 1\arcs, yielding an instrumental resolution of 4.8\,\AA\ FWHM, which is an instrumental dispersion of 108\,\kms, as measured from the night sky lines. For each galaxy we typically obtained a single 15 minute exposure along the major axis. In our survey the average image quality is 2\arcsec\ FWHM. 

In each exposure the spurious pixels are masked using a bad pixel map and Lacosmic\cite{2001PASP..113.1420V} for the cosmic rays. The frames are overscan and bias subtracted, and then flat fielded. The data is then interpolated in one step onto a frame linear in spatial direction and log-wavelength, using the wavelength solution from the nightly arc lamps and a spatial distortion map from a dedicated pinhole exposure. A variance (error) frame is propagated in the same way. During an exposure the effective aperture of the HET continually changes, which affects the throughput along the spatial direction. We correct for the throughput change using the apparent brightness of the sky lines in each science exposure. 

The central stellar absorption line kinematics are extracted by applying pPXF\cite{2004PASP..116..138C} with 120 stellar templates from the MILES library\cite{2006MNRAS.371..703S,2011A&A...532A..95F} to the observed spectra within a $1.5 \times 1$\arcs\ aperture centered on the brightest pixel in the slit. The stellar kinematics and their errors are determined using 200 Monte Carlo simulations per spectra. When present, emission lines are detected using GANDALF\cite{2006MNRAS.366.1151S}. Inside the central aperture the S/N is typically over 100, resulting in robust velocity dispersion measurements. We have 73 galaxies in common with the NYU-VAGC and we found excellent agreement with their dispersion measurements (6\% standard deviation) when measured inside a similar 3\arcsec\ diameter aperture. 

For the dynamical modeling of NGC\,1277, we obtained a separate 15 minute exposure with the 1\arcsec\ slit along the major axis in good weather and seeing conditions. This allowed us to confirm our earlier measurement and to probe the kinematics at larger radii. To ensure reliable extraction we combine spatial rows into bins with a minimum signal-to-noise of 35. The stellar kinematics are then extracted up to fourth Gauss-Hermite moment\cite{1993ApJ...407..525V} in each bin. In the outer bins the dispersion is too low for a reliable extraction of the Gauss-Hermite moment, and they are thus omitted from these fits. 

The positioning and PSF of the observations are critically important for obtaining a reliable dynamical black-hole-mass measurement. For instance, if the PSF is underestimated, the black hole mass would be underestimated as well. Because the  luminosity model (See next section) is based on the high resolution \emph{HST} image, it can be used to reconstruct the PSF and positioning of the slit \emph{a posteriori}. We iteratively fitted the reconstructed slit image to the MGE model convolved with a PSF consisting of multiple round Gaussians. We found that a two-component Gaussian PSF was required for an adequately reconstructed slit image. The two Gaussians in the best-fit PSF have dispersions of 0.62\arcsec\ and 2.8\arcsec, with a fraction of 66 and 34\%, respectively.  The FWHM of this two-component PSF is 1.6\arcsec. 
The best-fit single-Gauss model for the PSF has a dispersion of 0.78\arcsec, and is much more compact. Using this PSF in the dynamical modeling reduces the black hole mass by 6\%, which is well within our 1$\sigma$ confidence interval.

\section{\emph{HST}/ACS Photometry\\}
\label{sec:ACSphot}
     
There is archival imaging available of NGC\,1277 from the NASA/ESA \emph{Hubble Space Telescope}, which was retrieved from the Hubble Legacy Archive, which is a collaboration between the Space Telescope Science Institute (STScI/NASA), the Space Telescope European Coordinating Facility (ST-ECF/ESA) and the Canadian Astronomy Data Centre (CADC/NRC/CSA). These observations are associated with program GO:10546  (PI:Fabian) and are in the F550M (narrow V) and F625W (Sloan $r$) filters. From these high resolution images it is immediately clear that a superposition of multiple galaxies does not explain the high velocity dispersion. With Galfit\cite{2002AJ....124..266P} we construct a  Multi Gauss Expansion\cite{1992A&amp;A...253..366M} (MGE) model of  NGC\,1277 in a joint fit with neighboring galaxy NGC\,1278, a giant elliptical galaxy at an angular separation of 47\arcsec. In addition, all other sources in the frame are masked by generously large apertures and ignored in the fit. The resulting MGE is a two-dimensional representation of the light distribution of NGC1277, consisting of 10 co-axial Gaussian components (see Table~S\ref{tab:MGE}). For our photometric analysis, we used a reconstructed point spread function obtained with the TinyTim package\cite{1995ASPC...77..349K}, replicating the dither pattern of the separate exposures.

NGC\,1277 is a lenticular (S0) galaxy which is flattened, cuspy, and has a small central regular dust disk that is seen nearly edge-on: its semi-major axis length is 0.13\,kpc, and apparent axis ratio is 0.3. Assuming that the dust ring is in the equatorial plane of this disk galaxy, it suggests that the inclination of the galaxy is 75$^\circ$.
 
We also performed a single-component Sersic model fit with Galfit, in the same way as the MGE. The best-fitting single-component model has a Sersic index $n=2.2$, a circularized half-light radius of $R_e = 1.0$\,kpc, and a projected axis ratio of $b/a = 0.61$. The ellipse that contains half of the light has a major axis of $R_{1/2} = 1.2$\,kpc. The inclination of the dust-ring in combination with the projected axis ratio of the overall light distribution, under the assumption of oblateness, implies that the intrinsic short-to-long axis ratio is $\sim$0.4.

However, a single component Sersic model is not a good representation of the data, and we improved the photometric model by adding multiple components. A good fit is achieved when combining four Sersic components. These represent the following features: 1) a central cusp with $R_{1/2} < 100$~pc that accounts for 3\% of the total luminosity; 2) an inner round $n\simeq1$ component with $R_{1/2} = 0.3$ kpc, that accounts for 24\%; 3) an outer flatter $n\simeq1$ component with $R_{1/2} = 1.5$ kpc that accounts for 53\%; and 4) a rounder ($\epsilon \simeq 0.6$), low-$n$ halo component with $R_{1/2} = 4.4$ kpc that accounts for 20\%. None of these components has a high Sersic index---a classical bulge appears to be entirely absent. The total F550M magnitude, corrected for galactic extinction (0.545 mag) is 13.40 in the AB system, corresponding to an absolute V band magnitude of $-21.0$. 

\section{Schwarzschild Models\\}
\label{sec:dynmass}

Black hole masses are most often measured using Schwarzschild orbit-superposition models\cite{1979ApJ...232..236S} of the stellar kinematics. The triaxial Schwarzschild code used here is described in detail and the  recovery of the internal dynamical structure, or distribution function, intrinsic shape and black hole mass all validated in a series of papers\cite{2008MNRAS.385..647V,2008MNRAS.385..614V,2009MNRAS.398.1117V,2010MNRAS.401.1770V, 2012ApJ...753...79W}.

The apparent flatness and disky appearance of the galaxy indicates that the galaxy must be (nearly) axisymmetric and seen close to edge-on. The strong rotation observed around the short axis, indicates that this galaxy must be (nearly) oblate axisymmetric. In our luminosity model the flattest Gaussian has an axis ratio of 0.4, which indicates that this galaxy can not be seen much more face-on that than 70\dgr\ as it would otherwise be very elongated, or triaxial and intrinsically much flatter than it appears. Prolate shapes are ruled out due to strong rotation seen around the (apparent) short axis. 

We also assume that mass-to-light ratio of the stars is constant, which is justified by the lack of strong color gradients in the photometry.  The remaining variables are black hole mass, mass-to-light ratio and the dark halo. For the dark matter, we used a two-parameter spherical NFW halo\cite{1996ApJ...462..563N} with densities of $10^{-5}<\rho_0 <10$ \Msun\ pc$^{-3}$ and scale radii of $1<R_s<2000$ kpc. We also probed models without dark matter. In total there are thus only 4 free parameters in these models: black hole mass, mass-to-light ratio and two from the dark halo. We used iterative refinement to localize the minimum in parameter space and then found the location of the confidence intervals by iteratively doing small orthogonal steps around all reasonable models in all possible directions until convergence was reached. To explore parameter space, we ran over 600\,000 models on the THEO supercomputer of the Max Planck Institute for Astronomy, hosted at the Rechenzentrum of the Max Planck Society in Garching.

\begin{figure}
	\centering
	\includegraphics[width=8.3cm]{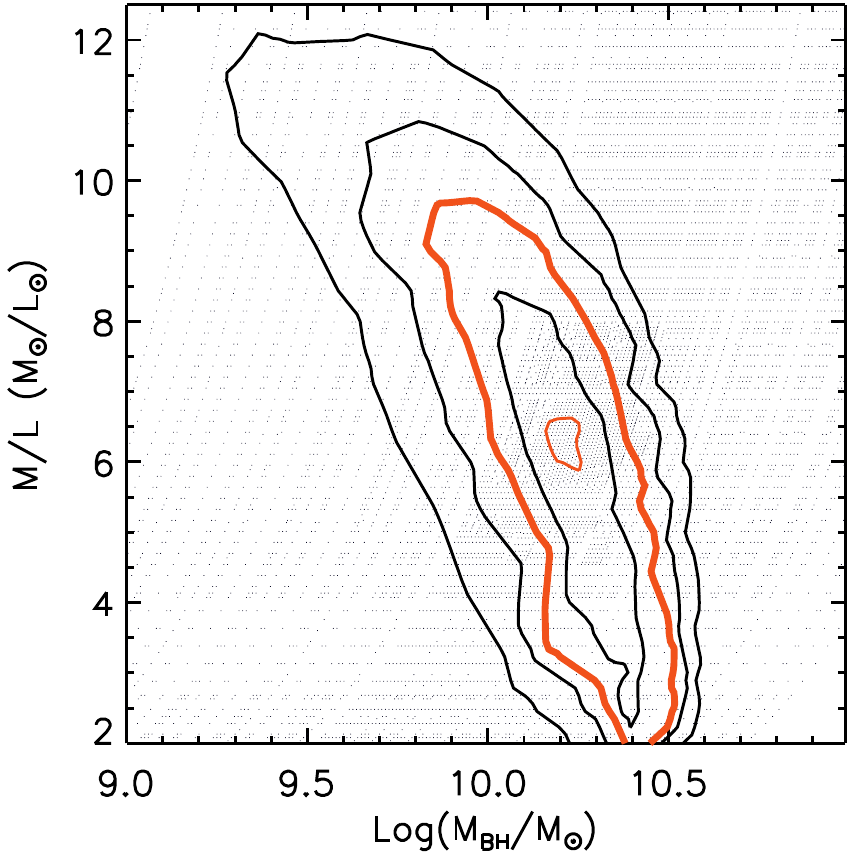}
\caption{Goodness-of-fit contours for 600\,000 dynamical models (black dots) of NGC\,1277 with different $V$-band mass-to-light ratios and black hole masses, after marginalizing over the dark halo parameters; i.e. for every combination of $M/L$ and \Mbh\ the best-fitting dark matter halo is chosen. The contours indicate 1$\sigma$ (red), 2$\sigma$, 3$\sigma$ (red, thick), 4$\sigma$ and 6$\sigma$ confidence levels for one degree of freedom.}
	\label{fig:N1277chi}
\end{figure}

\begin{figure}
	\centering
	\includegraphics[width=8.3cm]{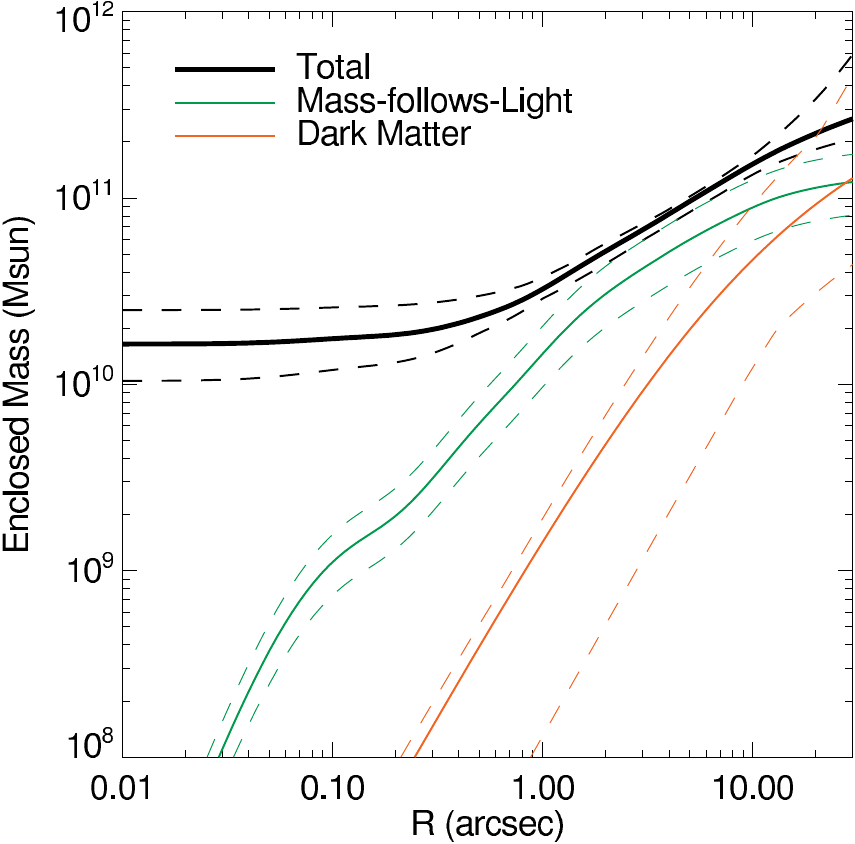}
\caption{Enclosed mass as function of radius as inferred from the dynamical models of NGC\,1277. The solid line indicated the best-fit model and the dashed lines the 68\% confidence interval. Inside 1.6\arcsec the black hole mass dominates, then the stars take over until 20\arcsec, where the dark matter density becomes larger than the stellar density. The dark matter content is not well constrained because the stellar kinematics only reach out to 20\arcsec.}
	\label{fig:enclosedmass}
\end{figure}

The resulting mass distribution has a stellar dynamical V-band mass-to-light ratio of $6.4^{+2.2}_{-0.8}$ \Msun/$L_{\odot_V}$ and a black hole mass of 17$\pm3\times 10^9$ \Msun\ (Fig.~S\ref{fig:N1277chi}).  The best-fitting NFW halo has $\rho_0=0.027$ \Msun\  pc$^{-3}$ and $R_s$=26 kpc, but the DM halo is only marginally constrained: the best-fitting model without a dark halo is within 3$\sigma$ with $\Delta\chi^2$ of only $6.3$. The confidence intervals of the dark matter content are shown in Fig.~S\ref{fig:enclosedmass}. The dark matter becomes the dominant component at 20\arcsec, which is at the limit of our long slit observations. The dark matter could thus be better constrained with kinematic tracers at large radii. The orbit distribution of the best-fit model is mildly anisotropic, $\sigma_r/ \sigma_\theta=0.8$, and is almost constant as a function of radius.

By comparing the total dynamical mass, derived from the kinematics, to the total spectrophotometric stellar mass, the stellar population can be constrained, because the total mass-follows-light mass is an upper limit to the total stellar content. From the dynamical model we found a mass-follows-light mass of $1.2\times 10^{11}$ \Msun\ which is in good agreement with $1.5\times 10^{11}$ \Msun\ based on spectral synthesis of the SDSS spectrum\cite{2005MNRAS.358..363C} with Bruzual \& Charlot stellar population models\cite{2003MNRAS.344.1000B} and a Chabrier\cite{2003PASP..115..763C} initial stellar mass function (IMF). This mass estimate is also in agreement with  the spectro-photometric mass\cite{2003MNRAS.341...33K} derived from  multi-band SDSS photometry. A Salpeter IMF would increase the  stellar mass-to-light by at least 60\% to 9.6 \MLsunV, which is strongly ruled out (Fig.~S\ref{fig:N1277chi}). 

\begin{figure}[t]
	\centering
	\includegraphics[width=8.3cm]{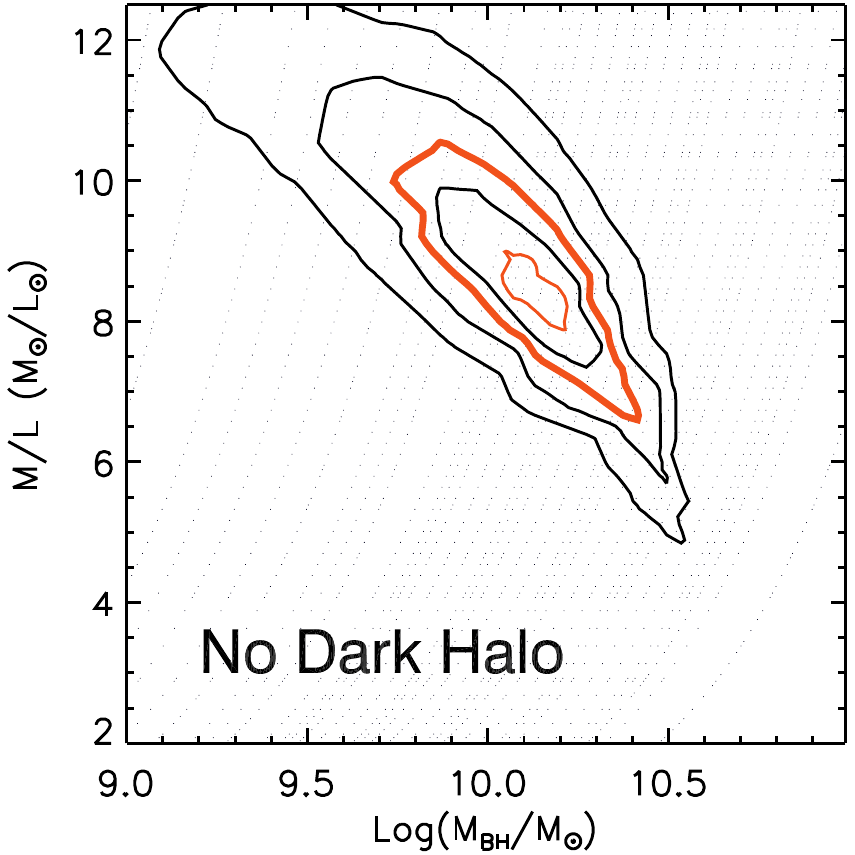}
\caption{Goodness-of-fit contours for the subset of dynamical models without dark matter  of NGC\,1277 (black dots) with different $V$-band mass-to-light ratios and black hole masses.  The contour levels are the same as in Fig.~S\ref{fig:N1277chi}. The best-fitting model without dark matter has $\Delta\chi^2$ offset of 6.3 (2.5$\sigma$) compared to the overall best-fit with dark matter. With this conservative choice of no-dark matter the black hole mass decreases by less than $2\sigma$.}
	\label{fig:N1277chi22_nodm}
\end{figure}

The best-fitting model is a good fit to the data. The model accurately reproduces the flat rotation curve and dispersion profile, including the central peak. To check if there was a better minimum at black hole mass below $10^8$~\Msun\ that was missed by our iterative search, we restarted the parameter search and enforced this upper limit. The resulting optimal model is a very poor fit to the kinematics and is strongly ruled out. To investigate the model dependence on the dark matter, we show the constraints on the $M/L$ and black hole mass for subset of models without dark matter in Fig.~S\ref{fig:N1277chi22_nodm}. This conservative estimate only lowers the measured black hole mass by less than $2\sigma$.  The high velocity dispersion peak can also not be caused by a nuclear stellar cluster as there is no strong colour gradient seen in either the SDSS or the \emph{HST} images. Future high-resolution spectroscopy is required to study exotic possibilities, such as an exponentially rising IMF. 
 
The black hole is easily resolved by these observations. The radius at which the total enclosed mass of stars is equal to the black hole mass is 1.6\arcsec, as shown in Fig.~S\ref{fig:enclosedmass}. To illustrate the effect of the black hole on the kinematics, we also show a model with a smaller $10^8$ \Msun\ black hole in Fig.~\ref{fig:NGC1277slit}. This model has the same dark halo and mass-to-light ratio as the best-fit model and it fails to reproduce the central dispersion profile. The HET long-slit kinematics can thus reliably resolve the black hole in NGC\,1277.\footnote{This document
has been typeset from a \TeX / \LaTeX\ file prepared by the author.}

\begin{table}[tb!]
\begin{center}
\begin{tabular}{ccc}
\hline \hline
$\rm m_{V} $  &  log $\sigma '$ & $q'$ \\ 
(mag)    &  (arcsec) &  \\ 
 \hline    
    19.079 &  -1.257 &  0.900 \\
    17.821 &  -0.634 &  0.832 \\
    16.650 &  -0.164 &  0.532 \\
    16.404 &   0.009 &  0.801 \\
    15.848 &   0.365 &  0.511 \\
    16.554 &   0.706 &  0.446 \\
    16.046 &   0.720 &  0.418 \\
    15.837 &   0.924 &  0.502 \\
    17.232 &   1.023 &  0.655 \\
    16.045 &   1.230 &  0.852 \\
\hline
\end{tabular}
\end{center}
\caption{Luminosity model of NGC 1277 composed of 10-Gaussians based on the F550M-band \emph{HST}/ACS image (GO:10546). The position angle of the galaxy is 92.7\dgr. Col.\ (1): Apparent AB magnitude, without a correction for galactic foreground extinction{\protect \cite{1998ApJ...500..525S}} (0.545 Mag). Col.\ (2): size along the major axis. Col.\ (3): flattening}

\label{tab:MGE}
\end{table}

\begin{table*}[t]
	\begin{center}
	\begin{tabular}{ccccccc}
	\hline \hline
	x$_b$ & x$_e$ & V & $\sigma$ & $h_3$ & $h_4$ \\
	(arcsec) & (arcsec) &  (\kms) & (\kms)  & & \\
	\hline
-22.87 & -13.94 & -205.47$\pm$30.25 &  52.43$\pm$66.41 &  - & - \\
-13.94 &  -9.71 & -248.56$\pm$16.49 & 109.19$\pm$35.82 &  - & - \\
 -9.71 &  -7.83 & -262.10$\pm$15.39 & 152.08$\pm$22.92 &  - & - \\
 -7.83 &  -6.42 & -283.81$\pm$13.32 & 170.37$\pm$21.11 &  - & - \\
 -6.42 &  -5.48 & -306.36$\pm$15.91 & 165.06$\pm$17.29 &  0.09$\pm$0.09 & 0.21$\pm$0.06 \\
 -5.48 &  -4.54 & -273.66$\pm$14.50 & 186.12$\pm$18.34 &  0.09$\pm$0.08 & 0.12$\pm$0.06 \\
 -4.54 &  -4.07 & -259.05$\pm$21.70 & 175.08$\pm$23.93 &  0.05$\pm$0.09 & 0.16$\pm$0.11 \\
 -4.07 &  -3.60 & -263.14$\pm$17.36 & 202.04$\pm$28.02 &  0.08$\pm$0.07 & 0.12$\pm$0.10 \\
 -3.60 &  -3.13 & -259.07$\pm$18.79 & 212.23$\pm$33.89 &  0.13$\pm$0.08 & 0.12$\pm$0.12 \\
 -3.13 &  -2.66 & -255.88$\pm$16.05 & 235.80$\pm$21.71 &  0.09$\pm$0.06 & 0.03$\pm$0.06 \\
 -2.66 &  -2.19 & -228.79$\pm$14.64 & 269.55$\pm$15.48 &  0.09$\pm$0.05 & 0.02$\pm$0.06 \\
 -2.19 &  -1.72 & -218.96$\pm$11.62 & 288.97$\pm$14.52 &  0.08$\pm$0.04 & 0.03$\pm$0.04 \\
 -1.72 &  -1.25 & -199.93$\pm$12.57 & 313.87$\pm$11.80 &  0.10$\pm$0.03 & 0.02$\pm$0.03 \\
 -1.25 &  -0.78 & -169.91$\pm$12.43 & 346.04$\pm$12.32 &  0.08$\pm$0.02 & 0.04$\pm$0.02 \\
 -0.78 &  -0.31 & -112.81$\pm$11.85 & 382.12$\pm$12.98 &  0.04$\pm$0.02 & 0.04$\pm$0.03 \\
 -0.31 &   0.16 &  -28.02$\pm$11.80 & 415.85$\pm$13.61 &  0.01$\pm$0.02 & 0.04$\pm$0.02 \\
  0.16 &   0.63 &   68.77$\pm$12.25 & 400.93$\pm$12.78 & -0.02$\pm$0.02 & 0.04$\pm$0.02 \\
  0.63 &   1.10 &  144.74$\pm$11.20 & 361.90$\pm$12.80 & -0.06$\pm$0.02 & 0.05$\pm$0.02 \\
  1.10 &   1.57 &  192.30$\pm$13.53 & 330.18$\pm$12.11 & -0.10$\pm$0.03 & 0.03$\pm$0.03 \\
  1.57 &   2.04 &  208.42$\pm$11.86 & 300.62$\pm$12.51 & -0.08$\pm$0.03 & 0.03$\pm$0.04 \\
  2.04 &   2.51 &  222.60$\pm$12.75 & 279.00$\pm$15.72 & -0.09$\pm$0.04 & 0.03$\pm$0.05 \\
  2.51 &   2.98 &  243.41$\pm$15.43 & 253.36$\pm$17.05 & -0.08$\pm$0.05 & 0.02$\pm$0.06 \\
  2.98 &   3.45 &  261.17$\pm$17.56 & 224.84$\pm$28.72 & -0.13$\pm$0.07 & 0.05$\pm$0.08 \\
  3.45 &   3.92 &  261.08$\pm$17.74 & 209.80$\pm$31.98 & -0.10$\pm$0.07 & 0.11$\pm$0.11 \\
  3.92 &   4.39 &  259.14$\pm$18.90 & 186.68$\pm$24.86 & -0.06$\pm$0.08 & 0.15$\pm$0.10 \\
  4.39 &   4.86 &  269.81$\pm$22.35 & 174.05$\pm$25.79 & -0.04$\pm$0.11 & 0.14$\pm$0.11 \\
  4.86 &   5.80 &  278.65$\pm$14.28 & 187.11$\pm$16.95 & -0.11$\pm$0.08 & 0.14$\pm$0.06 \\
  5.80 &   6.74 &  311.25$\pm$17.63 & 153.60$\pm$19.84 & -0.08$\pm$0.10 & 0.21$\pm$0.07 \\
  6.74 &   8.15 &  280.20$\pm$14.37 & 167.29$\pm$21.58 &  - & - \\
  8.15 &  10.50 &  257.91$\pm$14.26 & 146.68$\pm$23.17 &  - & - \\
 10.50 &  25.54 &  235.19$\pm$16.47 &  84.60$\pm$45.37 &  - & - \\
	 
	\hline
	\end{tabular}
		 
	\end{center}
	\caption{The line-of-sight stellar kinematics of NGC\,1277 obtained through a 1\arcsec\ slit with the LRS{\protect \cite{1998SPIE.3355..375H}} on the HET. The position angle of the slit is 95 degrees North-East. The instrumental resolution is 108\,\kms\ as determined from the night sky lines.  Col.\ (1,2): Begin and end of each spatial bin along the slit. Col.\ (3): Mean  velocity. Col.\ (4): Velocity dispersion. Col.\ (5,6): Gauss-Hermite Moments $h_3$ and $h_4$.}
	\label{tab:NGC1277kin}
\end{table*}

\section*{Additional References}

\end{document}